\documentclass[twocolumn]{aastex63}

\usepackage{lineno}
\usepackage[figuresright]{rotating}

\received{***, ****}
\revised{***, ****}
\accepted{***, ****}
\submitjournal{ApJ}

\begin{document}
\title{Comparison of Ion-Proton Differential Speed between ICMEs and Solar Wind near 1 au}

\author{Xuechao Zhang}
\affiliation{Institute of Frontier and Interdisciplinary Science, Shandong University, Qingdao, Shandong 266237, China}
\affiliation{Shandong Provincial Key Laboratory of Optical Astronomy and Solar-Terrestrial Environment, and Institute of Space Sciences, Shandong University, Weihai, Shandong 264209, China}

\correspondingauthor{Hongqiang Song}
\email{hqsong@sdu.edu.cn}

\author{Hongqiang Song}
\affiliation{Shandong Provincial Key Laboratory of Optical Astronomy and Solar-Terrestrial Environment, and Institute of Space Sciences, Shandong University, Weihai, Shandong 264209, China}

\author{Chengxiao Zhang}
\affiliation{Shandong Provincial Key Laboratory of Optical Astronomy and Solar-Terrestrial Environment, and Institute of Space Sciences, Shandong University, Weihai, Shandong 264209, China}

\author{Hui Fu}
\affiliation{Shandong Provincial Key Laboratory of Optical Astronomy and Solar-Terrestrial Environment, and Institute of Space Sciences, Shandong University, Weihai, Shandong 264209, China}

\author{Leping Li}
\affiliation{National Astronomical Observatories, Chinese Academy of Sciences, Beijing, 100101, China}

\author{Jinrong Li}
\affiliation{Shandong Provincial Key Laboratory of Optical Astronomy and Solar-Terrestrial Environment, and Institute of Space Sciences, Shandong University, Weihai, Shandong 264209, China}

\author{Xiaoqian Wang}
\affiliation{Shandong Provincial Key Laboratory of Optical Astronomy and Solar-Terrestrial Environment, and Institute of Space Sciences, Shandong University, Weihai, Shandong 264209, China}

\author{Rui Wang}
\affiliation{Shandong Provincial Key Laboratory of Optical Astronomy and Solar-Terrestrial Environment, and Institute of Space Sciences, Shandong University, Weihai, Shandong 264209, China}

\author{Yao Chen}
\affiliation{Shandong Provincial Key Laboratory of Optical Astronomy and Solar-Terrestrial Environment, and Institute of Space Sciences, Shandong University, Weihai, Shandong 264209, China}
\affiliation{Institute of Frontier and Interdisciplinary Science, Shandong University, Qingdao, Shandong 266237, China}

\begin{abstract}
The elemental abundance of ICMEs and solar wind near 1 au is often adopted to represent the abundance in the corresponding coronal sources. However, the absolute abundance of heavy ions (relative to hydrogen) near 1 au might be different from the coronal abundance due to the ion-proton differential speed ($V_{ip}$). To illustrate the $V_{ip}$ characteristics and explore whether it influences the absolute abundance analysis for ICMEs and solar wind, we perform a statistical study on the $V_{ip}$ for He$^{2+}$, C$^{5+}$, O$^{6+}$, and Fe$^{10+}$ in both ICMEs and solar wind based on measurements of Advanced Composition Explorer. The results show that the $V_{ip}$ is negligible within ICMEs and slow solar wind ($<$ 400 km s$^{-1}$), while obvious in the intermediate (400 -- 600 km s$^{-1}$) and fast wind ($>$ 600 km s$^{-1}$). Previous studies showed that the $V_{ip}$ in ICMEs keeps negligible during propagation from 0.3 to 5 au, but in solar wind it increases with the decreasing heliocentric distance. Therefore, it might be questionable to infer the absolute abundance of coronal sources through in-situ abundance near 1 au for solar wind. Fortunately, the ion-oxygen (O$^{6+}$) differential speed ($V_{io}$) is negligible for He$^{2+}$, C$^{5+}$, and Fe$^{10+}$ within both ICMEs and solar wind, and previous studies suggested that the $V_{io}$ does not vary significantly with the heliocentric distance. This indicates that various heavy ions always flow at the same bulk speed and their relative abundance (relative to oxygen) near 1 au can represent the coronal abundance for both ICMEs and solar wind.
\end{abstract}


\section{Introduction}
Interplanetary coronal mass ejections (ICMEs) are the counterpart of coronal mass ejections \citep[CMEs,][]{forbes06,chenpengfei11,webb12} in the interplanetary space \citep{gopalswamy06,jianlan06a,manchester17}. CMEs are often associated with eruption of hot channels \citep[e.g.,][]{zhangjie12,chengxin13b,song14a} or solar filaments \citep[e.g.,][]{song18a,chengxin20,wangxinyue23}. Solar wind is the continuous flow of charged particles from the Sun and usually is divided into slow and fast wind \citep{mccomas03,ofman10,vidotto21}. Both ICMEs and solar wind play significant roles in space weather \citep[e.g.,][]{zhangjie03,zhangjie07,xumengjiao19} and are important research topics in solar and space physics.

One important research field about ICMEs and solar wind is their elemental abundance of heavy ions, including the absolute abundance (relative to hydrogen) and relative abundance (generally relative to oxygen) \citep[e.g.,][]{lepri13,zurbuchen16,owens18,song20b,song21b}. We can acquire the elemental abundance of ICMEs directly by means of the measurement of spacecraft near 1 au, with which to infer their plasma origin. For example, an ICME induced by the filament eruption arrived at the vicinity of the Earth on 1998 May 1. Researchers analyzed the elemental abundance of the filament such as Fe/O, Mg/O, and Si/O. They found that these abundances are close to the corresponding photospheric values \citep{song17a,lepri21}, supporting that the filament plasma originates from the chromosphere/photosphere \citep{wangjincheng18,wangjincheng19} instead of corona.

For the solar wind, researchers usually trace the parcels measured near 1 au back to the 2.5 $R_\odot$ source surface via the ballistic mapping assuming a constant bulk speed of solar wind, then map them back to the magnetic footpoints on the solar surface \citep{neugebauer98,neugebauer02,neugebauer04} through the potential field source surface model \citep{altschuler69,schatten69}. With this two-step mapping procedure, \cite{zhaoliang17a} categorized the solar wind into different types according to their coronal source regions, then examined and compared their relative abundances. \cite{fuhui18} performed a statistical comparison on the absolute abundance of helium and the alpha-proton differential streaming ($V_{\alpha p}$) for solar wind from coronal holes, active regions, and quiet-Sun regions.

The $V_{\alpha p}$ in solar wind was reported for the first time in 1970s \citep{robbins70,formisano70}. Since then, it has been investigated intensively by analyzing different spacecraft observations at various heliocentric distances from $<$0.1 to $\sim$5.4 au \citep{marsch82b,neugebauer96,durovcova17,alterman18,mostafavi22,mostafavi24,ranhao24}. The observations showed that the $V_{\alpha p}$ in solar wind decreased continually with increasing distance from the Sun until $\sim$4.2 au at which no significant differential streaming was detected \citep{neugebauer96,reisenfeld01}. At various distances, the $V_{\alpha p}$ direction is aligned with the interplanetary magnetic field (IMF) \citep{marsch82b,kasper08} and its magnitude keeps comparable to or lower than the local Alfv\'{e}n speed ($V_{A}$) \citep{marsch82b,neugebauer94,neugebauer96,mostafavi22,mostafavi24}. Besides, previous studies also demonstrated that the $V_{\alpha p}$ is proportional to the solar wind speed. It is very small for the slow solar wind ($<$ 400 km s$^{-1}$), while in the fast wind ($>$ 500 km s$^{-1}$), it can be as large as 150 km s$^{-1}$ at 0.3 au and decreases to $\sim$40 km s$^{-1}$ near 1 au \citep{marsch82b}. Recent observations of Parker Solar Probe \citep[PSP,][]{fox16} showed that the $V_{\alpha p}$ in slow wind ($<$ 400 km s$^{-1}$) reaches $\sim$60 km s$^{-1}$ near 0.1 au, while it is still negligible around 0.3 au \citep{marsch82b}. For the wind at speed of 500 -- 600 km s$^{-1}$, the $V_{\alpha p}$ is $\sim$160 km s$^{-1}$ and $\sim$110 km s$^{-1}$ around 0.1 au and 0.3 au \citep{mostafavi22}, respectively.

Several mechanisms have been proposed to explain the preferential acceleration and heating of alpha particles, such as the resonant absorption of ion--cyclotron waves \citep{kasper13} and the stochastic heating caused by low-frequency Alfv\'{e}n wave turbulence \citep{chandran10}. \cite{verscharen13b} discussed the role of the Alfv\'{e}nic drift instability for the $V_{\alpha p}$ evolution in the solar wind. In the meantime, the wave-particle interaction driven by alpha-proton instabilities \citep{gary00} and the Coulomb collisions \citep{kasper08} reduce the $V_{\alpha p}$ down to or below the local $V_{A}$. The kinetic microinstability can also limit the temperature anisotropy of protons and alpha particles \citep{maruca12}. The collisional age ($A_{c}$) is the ratio of the estimated transit time of the solar wind based on the local speed to the time between the local Coulomb collisions. It is a useful proxy to evaluate the importance of collisions. When $A_{c} \ll 1$ the plasma is collisionless; when $A_{c} \gg 1$ the plasma is collisionally old and Coulomb collisions reduce the nonequilibrium features greatly \citep{kasper08,chhiber16,durovcova17}.

So far, the studies of differential streaming are mainly centered on the $V_{\alpha p}$. It is suggested that the kinetic properties of heavy ions behave in a well organized way under most solar wind flow conditions \citep{vonSteiger95}. The observations of ISEE-3 around the maximum of solar cycle 21 showed that iron and silicon ions tend to flow slower than alpha particles by $\sim$10--20 km s$^{-1}$ in the fast wind near 1 au, while no significant differential streaming exists in the slow wind \citep{schmid87,bochsler89}. SOHO observations in 1996 (solar minimum) displayed that iron and silicon ions tend to lag behind oxygen ions with a speed difference of $\sim$20 km s$^{-1}$ at 500 km s$^{-1}$ \citep{hefti98}. The Ulysses observations, covering a time range from 1991 to 1998 and a heliocentric distance from 1.8 to 5.4 au, showed that all heavy ion species flow with the equal bulk speed in both slow and fast winds \citep{vonSteiger95,vonSteiger06}, \textit{i.e.,} the isotachic property is obeyed better than the ISEE-3 and SOHO measurements. \cite{berger11} analyzed the differential streaming between 44 heavy ions and proton ($V_{ip}$) for two high-speed streams in early 2008 (solar minimum), and found that the $V_{ip}$ of 35 ion species lies within the range of $\pm$0.15 $V_{A}$ around 0.55 $V_{A}$, also indicating that different heavy ions flow at the same bulk speed.

The obvious $V_{ip}$ (including $V_{\alpha p}$) in the solar wind indicates that the absolute elemental abundance of heavy ions near 1 au might be different from the coronal abundance. To keep the flux conservation of various particle species, the number-density ratio of heavy ions and protons (the absolute abundance) is related to their bulk-speed ratio under a spherical geometry and steady state. Therefore, the abundance of solar wind varies as the bulk-speed ratio changes due to the $V_{ip}$ variation along the heliocentric distance. Besides, the abundance of coronal sources can vary with time \citep{widing01,baker13}, and the measured heavy ions and protons near 1 au are released at different times due to the $V_{ip}$, which also results in the abundance difference between the corona and 1 au. While heavy ions of solar wind might have the identical relative elemental abundance near the Sun and Earth.

We anticipate that both the absolute and relative elemental abundances of ICMEs remain unchanged during their propagation to 1 au, as previous study demonstrated that the $V_{\alpha p}$ is negligible within ICMEs from 0.3 au to 5 au \citep{liuying06} and recent study also suggested that the $V_{\alpha p}$ within ICMEs should be equal to zero near the Sun \citep{durovcova17}. To support the speculation, it is necessary to conduct a systematic and comprehensive study on both the $V_{ip}$ and $V_{io}$ (ion-oxygen differential streaming) within ICMEs, which is our motivation to conduct the current study. The paper is organized as follows. Section 2 introduces the instruments and methods. Section 3 presents the observations and results. Discussions are given in Section 4 that is followed by a summary in the final section.

\section{Instruments and Methods}
\subsection{Instruments}
All of the data used in this paper are from the Advanced Composition Explorer \citep[ACE,][]{stone98} spacecraft located at L1 point since 1997, and are available at the ACE science center\footnote{http://www.srl.caltech.edu/ACE/ASC/level2/index.html}. The proton velocity is provided by the Solar Wind Electron Proton and Alpha Monitor \citep[SWEPAM,][]{mccomas98}. SWEPAM observes plasma with independent electron and ion instruments by means of electrostatic analyzers, and can provide the full electron and ion distribution functions. The 1-D bulk speed of heavy ions (He$^{2+}$, C$^{5+}$, O$^{6+}$, and Fe$^{10+}$) are provided by the Solar Wind Ion Composition Spectrometer \citep[SWICS,][]{gloeckler98}. SWICS is a time-of-flight mass spectrometer and can separate the ions by their mass, charge, and energy through a triple coincidence measurement \citep[e.g.,][]{gilbert12}. Here we use the updated SWICS 1.1 level 2 data that have a better accuracy \citep{shearer14}. The IMF vector are provided by the ACE Magnetometer \citep[MAG,][]{smith98}. In this paper, the data with 1-h time resolution are used for the three payloads.

\subsection{Methods}
To analyze the alpha-proton differential flow, some previous studies used the speed difference ($V_{\alpha p}=|\mathbf{V_{\alpha}}|-|\mathbf{V_{p}}|$) directly \citep{asbridge76,neugebauer94,kasper08}. However, the directions of proton and alpha velocities are usually different, thus $|\mathbf{V_{\alpha}}|-|\mathbf{V_{p}}|=0$ does not mean no differential streaming \citep{durovcova17}. For this reason, some recent studies analyzed the $V_{\alpha p}$ using Equation (1) \citep{durovcova17,mostafavi22,mostafavi24,ranhao24}.
\begin{equation}
\rm V_{\alpha p}=sign(\left|\mathbf{{V}_{\alpha}}\right|-\left|\mathbf{{V}_{p}}\right|) \cdot \left|\mathbf{{V}_{\alpha}} - \mathbf{{V}_{p}}\right|
\end{equation}
Note that the bold symbols denote the vectors and the unbold symbols, the vector magnitudes.

As SWICS only provides the 1-D bulk speed of heavy ions, which can be taken as their radial speed, and the drift velocities are parallel to the magnetic field, we calculate $V_{ip}$ along the field using Equation (2) following previous studies \citep{reisenfeld01,fuhui18}
\begin{equation}
\rm V_{ip}=\frac{V_{ri}-V_{rp}}{cos\theta}
\end{equation}
where $V_{ri}$ and $V_{rp}$ are the radial speed of heavy ions and protons, respectively, and $\theta$ represents the angle between the magnetic field and the radial vector. To reduce the uncertainty, observations with $\theta$ greater than 72.5 degrees are discarded as done in previous studies \citep{reisenfeld01,fuhui18}.

As mentioned above, the $V_{ip}$ (including $V_{\alpha p}$) stays comparable to or lower then the local $V_{A}$. To further confirm this point in both ICMEs and solar wind, we also analyze the $(V_{ip}/V_{A}) \cdot 100\%$ . The $V_{A}$ is calculated according to Equation (3)
\begin{equation}
\rm V_A=\frac{B}{\sqrt{\mu_0\left(N_p m_p+N_\alpha m_\alpha\right)}}
\end{equation}
where B is the IMF magnitude, $\mu_{0}$ is the vacuum permeability, $N_p$ ($N_\alpha$) and $m_p$ ($m_\alpha$) are the number density and mass of the proton (alpha particle), respectively.

As mentioned above, we use the data with 1-h resolution in the current work. \cite{berger11} calculated the $|V_{\alpha p}|/V_{A}$ with time resolution of 12 min, which is similar to the result at 92-s resolution \citep{kasper08}. They redid the analysis at 1-h cadence and found a decrease of the $|V_{ip}|$ compared to the 12-min result \citep{berger11}, thus one should pay attention to both the calculation method and the time resolution when analyzing the differential speed.

\section{Observations and Results}
\subsection{ICME Catalog}
To show the $V_{ip}$ of ICMEs and solar wind separately, we need find out ICMEs first from background solar wind. Researchers identify ICMEs according to several typical features, including enhanced magnetic-field magnitude, smoothly changing magnetic-field direction, lower proton temperature, and lower plasma $\beta$ etc \citep{wucc11}. Several ICME catalogs are available online based on observations of different spacecraft \citep{richardson10,chiyutian16,nieves18,jianlan18}. Each catalog provides the start and end time of the ICME ejecta, which are used to analyze the $V_{ip}$ within ICMEs. In this paper, we use the catalog of Richardson and Cane \citep{richardson04,richardson10} based on ACE measurements\footnote{https://izw1.caltech.edu/ACE/ASC/DATA/level3/icmetable2.html}, as the heavy-ion data are from ACE/SWICS. The SWICS team provides the optimal data from 1998 February to 2011 August, during which the catalog lists 319 ICMEs in total \citep[e.g.,][]{fengxuedong18,song21a,lijinrong23}.

\subsection{Distribution of the $V_{ip}$}
As mentioned, the $V_{ip}$ is correlated with solar wind speed and it should be analyzed in different speed ranges. Following previous studies \citep[e.g.,][]{marsch82b,durovcova17}, we divide solar wind into three subsets based on their bulk speed, i.e., slow ($<$400 km s$^{-1}$), intermediate (between 400 and 600 km s$^{-1}$), and fast ($>$600 km s$^{-1}$) winds. Likewise, ICMEs are also categorized into slow, intermediate, and fast groups according to the average speed of each event \citep{durovcova17}.

Top panels in Figure 1 present the probability distributions of the alpha-proton differential speed ($V_{\alpha p}$) within both ICMEs (red) and solar wind (black). The data point number (N), mean, and median are shown in each panel. The positive $V_{\alpha p}$ value means that alpha particles move faster than protons, while the negative value indicates that protons overtake alpha particles. Our studies show that the speed difference is low in the slow wind with a mean about 8 km s$^{-1}$ as shown in Panel (a), while it obviously increases to $\sim$30 km s$^{-1}$ in the intermediate wind and $\sim$46 km s$^{-1}$ in the fast wind. Panels (b) and (c) show a preferred positive $V_{\alpha p}$ in solar wind with speed beyond 400 km s$^{-1}$. On the contrary, the speed difference keeps negligible in all ICMEs, regardless of their speeds, which agrees with previous results \citep{durovcova17}.

Bottom panels in Figure 1 shows the probability distributions of the alpha-proton differential speed in percent of local $V_{A}$. The ratios are generally below 1 for various solar wind velocities, which is consistent with the WIND observations \citep[e.g.,][]{durovcova17,fuhui18} at 1 au. The measurements of PSP \citep{mostafavi22}, Helios \citep{marsch82b}, and Ulysses \citep{neugebauer96} demonstrated that the $V_{\alpha p}$ is smaller than local $V_{A}$ at any heliocentric distance from below 0.1 au to beyond 5 au.

Figures 2--4, with the same format as Figure 1, present the probability distributions of the ion-proton differential speeds and the differential speeds in percent of local $V_{A}$ for carbon-proton (C$^{5+}$, $V_{cp}$), oxygen-proton (O$^{6+}$, $V_{op}$), and iron-proton (Fe$^{10+}$, $V_{Fep}$), respectively. All of them exhibit the similar distribution characteristics and average values compared with the $V_{\alpha p}$ in Figure 1. The $V_{ip}$ is negligible for all these heavy ions within ICMEs. This suggests that the kinetic properties of heavy ions also behave in a well organized way within ICMEs, similar to the solar wind \citep{vonSteiger95}. To further demonstrate this point, we show the speed difference among heavy ions ($V_{io}=V_{i}-V_{o}$) in next subsection.

\subsection{Distribution of the $V_{io}$}
Panel (a) in Figure 5 shows the probability distributions of the alpha-oxygen differential speed ($V_{\alpha o}$) within ICMEs (red) and solar wind (black). The data point number (N), mean and median are displayed in the panel. Here they are not divided into three subsets according to their speed, as the $V_{\alpha o}$ is negligible in all speed conditions. The means are -1.4 and 0.24 km s$^{-1}$ (almost equal to zero) in ICMEs and solar wind, respectively. Their medians are also very close to zero, thus they flow at the same radial speed as expected \citep{vonSteiger06}. Panel (b) presents the correlogram of the radial speeds of alpha particles versus O$^{6+}$ in ICMEs. The Pearson correlation coefficient (cc) is 0.9977, and the slope of linear fitting (k) is 0.9777. Panel (c) shows the same correlogram but for solar wind, which also exhibits high correlation between the two ion species. These clearly demonstrate that the speeds of alpha particles and O$^{6+}$ are equal over the full range of 300-1000 km/s flow, agreeing with previous studies \citep{vonSteiger95,vonSteiger06}.

Panels (a) in Figures 6 and 7, with the same format as Figure 5(a), present the probability distributions of the carbon-oxygen and iron-oxygen differential speeds ($V_{co}$ and $V_{Feo}$), respectively. Both means and medians of ICMEs and solar wind are close to zero. The slopes of speed fitting for the ion pairs are investigated, and all slopes are almost equal to unity. These indicate that the C$^{5+}$ and Fe$^{10+}$ also flow at the same speed with O$^{6+}$. Alpha particles, C$^{5+}$, O$^{6+}$, and Fe$^{10+}$ possess different mass, charge, as well as mass-per-charge, but flow at the same bulk speed. This implies that all heavy ions have the identical speed within both ICMEs and solar wind \citep{berger11}.

\subsection{Solar Cycle Dependence of the $V_{ip}$}
It has been noted the $V_{\alpha p}$ is different between solar minimum and maximum in both fast and intermediate solar wind near 1 au \citep{durovcova17}, while no obvious variation for slow wind. Our study shows that the ion-proton speed difference is negligible in both slow wind and ICMEs near 1 au. Therefore, it is only necessary to investigate whether the solar cycle dependence of $V_{ip}$ exists in the fast and intermediate wind. To explore the solar cycle dependence, Panels (a)--(d) in Figure 8 displays the yearly average values of $V_{\alpha p}$, $V_{cp}$, $V_{op}$, and $V_{Fep}$ from 1998 to 2011 sequentially with green (red) denoting the fast (intermediate) wind. The bars represent the standard deviations in each year. To reduce the influence of overlap, the dashed and solid lines are used to represent the bars of fast and intermediate winds, respectively.

In order to quantify the correlation relation between $V_{ip}$ and solar activity level, we calculate the Spearman correlation coefficients (cc) between yearly ion-proton differential speeds ($V_{\alpha p}$, $V_{cp}$, $V_{op}$, and $V_{Fep}$) and the yearly sunspot numbers (not shown), and present them in each panel. The results show that all the correlation coefficients are beyond 0.8 in both the fast and intermediate solar winds. This demonstrates that the $V_{ip}$ in solar wind possesses the obvious solar cycle dependence near 1 au.

\section{Discussions}
As mentioned, researchers have found the dependence of the differential streaming on heliocentric distance in fast solar wind. For slow wind, earlier study showed that the $V_{\alpha p}$ is very small and it undergoes a negligible evolution from 0.3 to 1 au according to the observations of Helios \citep{marsch82b}. While recent observations of PSP \citep{fox16} demonstrated that the dependence of $V_{\alpha p}$ on distance is valid for slow wind inside 0.1 au, where the $V_{\alpha p}$ is larger and noticeable \citep{mostafavi22}. Therefore, we suggest that the absolute elemental abundances of both slow and fast solar winds near 1 au might be different from those in coronal sources.

Previous studies reported a correlation between the $V_{\alpha p}$ and the ratio of the Coulomb collision time to the solar wind expansion time \citep{neugebauer76,marsch82b,neugebauer94}. When the ratio is less than $\sim$10, Coulomb collisions limit the $V_{\alpha p}$ \citep{neugebauer94}. Such conditions occur in ICMEs, where the $V_{\alpha p}$ is reduced by strong Coulomb collisions \citep{liuying06}. A recent survey with WIND measurements \citep{durovcova17} and our current study with ACE data confirmed that the $V_{\alpha p}$ is negligible in ICMEs near 1 au. The $V_{\alpha p}$ in ICMEs keeps negligible from 0.3 au to 5 au \citep{liuying06}, as the Coulomb collisions are stronger inside 1 au. Besides, \cite{durovcova17} reported that the $V_{\alpha p}$ within ICMEs does not exhibit any clear dependence on the collisional age and the ICME speed. They also suggested that the $V_{\alpha p}$ did not undergo an obvious evolution along the ICME path from the Sun to 1 au through comparing the 2D histogram of collisional age versus $V_{\alpha p}$. This means the absolute elemental abundance of ICMEs might keep unchanged during propagation, and the in-situ values near 1 au can represent the abundance of ICMEs when they are in the corona.

When solar wind or ICMEs propagate outward, their electron density decreases rapidly with the radial distance. This leads to the freeze-in of ionic charge states before solar wind and ICMEs enter into the interplanetary space \citep{owocki83}. Therefore, the ionic charge states near 1 au are often adopted to infer the evolution processes of solar wind and ICMEs in the corona. Usually researchers use the charge state ratios of different oxygen ions (e.g., O$^{7+}$/O$^{6+}$) or different carbon ions (e.g., C$^{6+}$/C$^{5+}$), as well as the average charge state of iron to infer the coronal temperature. The analyses of charge states and relative abundances of heavy ions should not be influenced by the differential streaming, as both previous and our current studies suggest that all heavy ions flow with the same bulk speed within both ICMEs and solar wind.

\section{Summary}
In this paper, we first conducted a statistical comparison of the ion-proton differential speed ($V_{ip}$) between ICMEs and solar wind with ACE data. The results showed that the $V_{ip}$ is negligible in ICMEs and slow wind ($<$400 km s$^{-1}$) but still obvious in intermediate and fast winds near 1 au. Given that the $V_{\alpha p}$ increases with the decreasing heliocentric distance and it is also obvious near the Sun in slow wind \citep{mostafavi22}, we suggest that the absolute elemental abundance near 1 au might be different from the coronal abundance for all types of solar winds. However, the absolute abundance of heavy ions within ICMEs might keep unchanged during their propagation as their alpha-proton differential speed is negligible and does not vary significantly from $\sim$0.3 au to 1 au \citep{liuying06,durovcova17}, which needs to be examined further by observations of PSP \citep{fox16} and Solar Orbiter \citep{muller20} in future.

Then we analyzed the ion-oxygen differential speed ($V_{io}$) of ICMEs and solar wind. The results showed that the $V_{io}$ is negligible within ICMEs and all types of solar winds, which means that the relative elemental abundance near 1 au could represent the abundance of ICMEs and solar wind when they are in the corona. All heavy ions flowing at the same bulk speed also implies that the charge state ratio or average charge state of heavy ions keep unchanged after freeze-in in the corona. This allows us to infer the freeze-in process of ionic charge states near the Sun based on in-situ observations near 1 au. At last, our analysis demonstrated that the $V_{ip}$ in both the intermediate and fast solar winds exhibits an obvious solar cycle dependence.

\acknowledgments We thank the anonymous referee for the comments and suggestions that helped to improve the original manuscript greatly. We are grateful to Drs. Bo Li (Shandong University), Jiansen He (Peking University), Pengfei Chen (Nanjing University), Ying D. Liu (National Space Science Center, CAS), Jia Huang (University of California, Berkeley), and Liang Zhao (University of Michigan) for their helpful discussions. This work is supported by the Strategic Priority Research Program of the Chinese Academy of Sciences (Grant No. XDB0560000), the NSFC grants 12373062, 11973031, and 12073042, as well as the National Key R\&D Program of China 2022YFF0503003 (2022YFF0503000). The authors acknowledge the use of data from the ACE science center.

\clearpage

\begin{figure*}[htb!]
\epsscale{1.0} \plotone{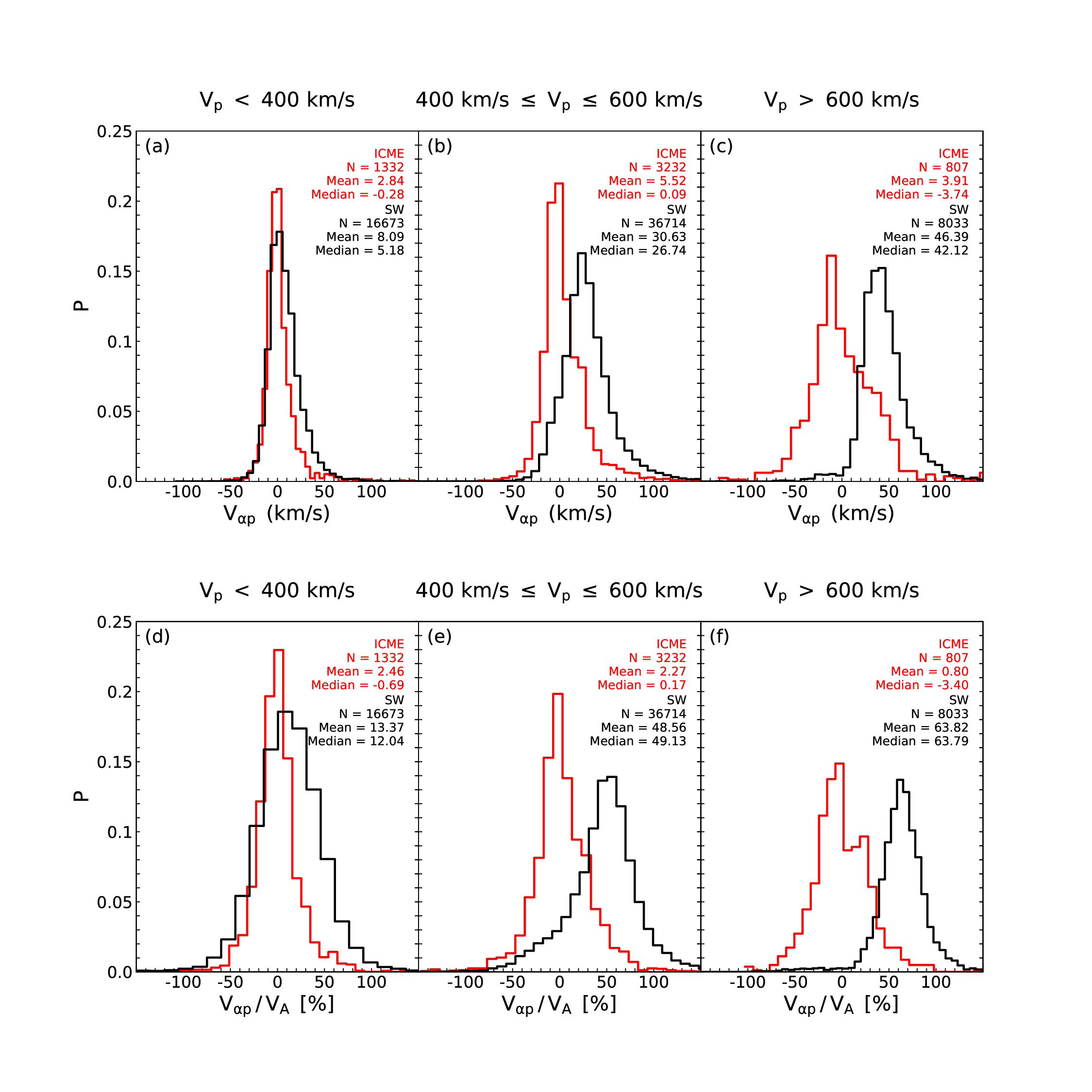} \caption{(Top) Probability distributions of the alpha-proton differential speed ($V_{\alpha p}$) for slow- (a), intermediate- (b), and fast- (c) ICMEs (red) and solar wind (black). (Bottom) Probability distributions of the alpha-proton differential speed normalized to local Alfv\'{e}n speed ($(V_{\alpha p}/V_{A}) \cdot 100\%$) correspondingly. \label{Figure 1}}
\end{figure*}

\begin{figure*}[htb!]
\epsscale{1.0} \plotone{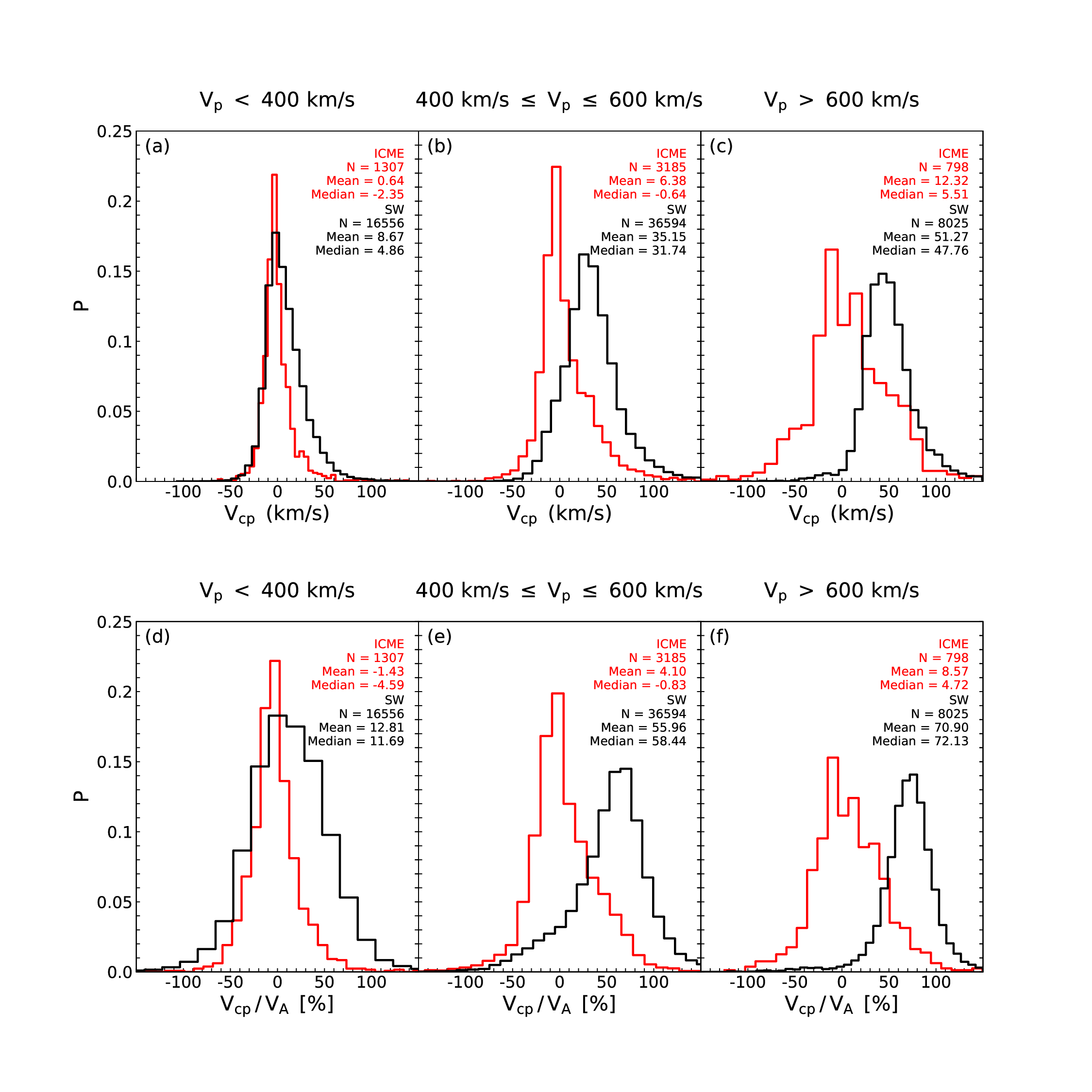} \caption{Same as Figure 1 but for the C$^{5+}$-proton differential speed ($V_{cp}$). \label{Figure 2}}
\end{figure*}

\begin{figure*}[htb!]
\epsscale{1.0} \plotone{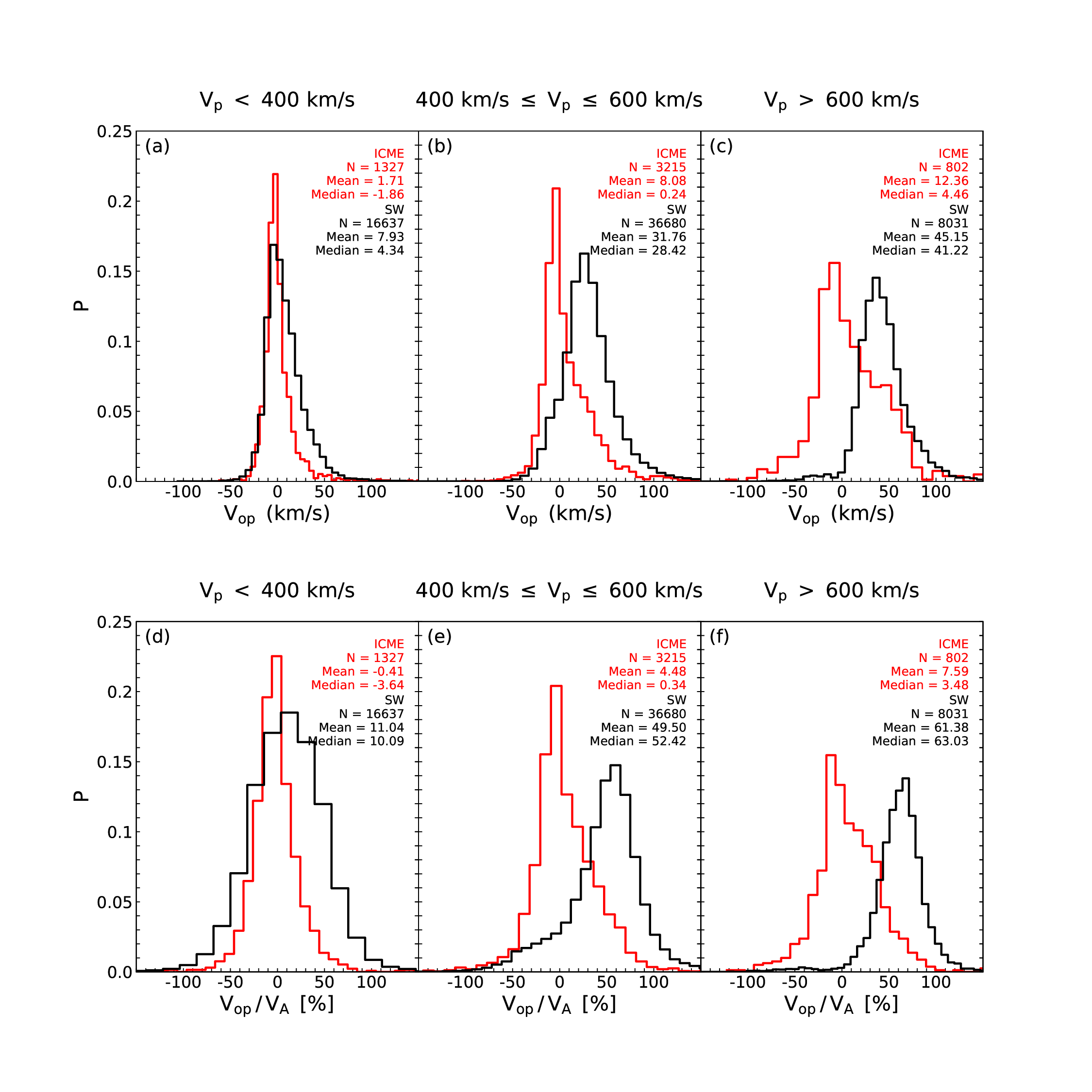} \caption{Same as Figure 1 but for the O$^{6+}$-proton differential speed ($V_{op}$). \label{Figure 3}}
\end{figure*}

\begin{figure*}[htb!]
\epsscale{1.0} \plotone{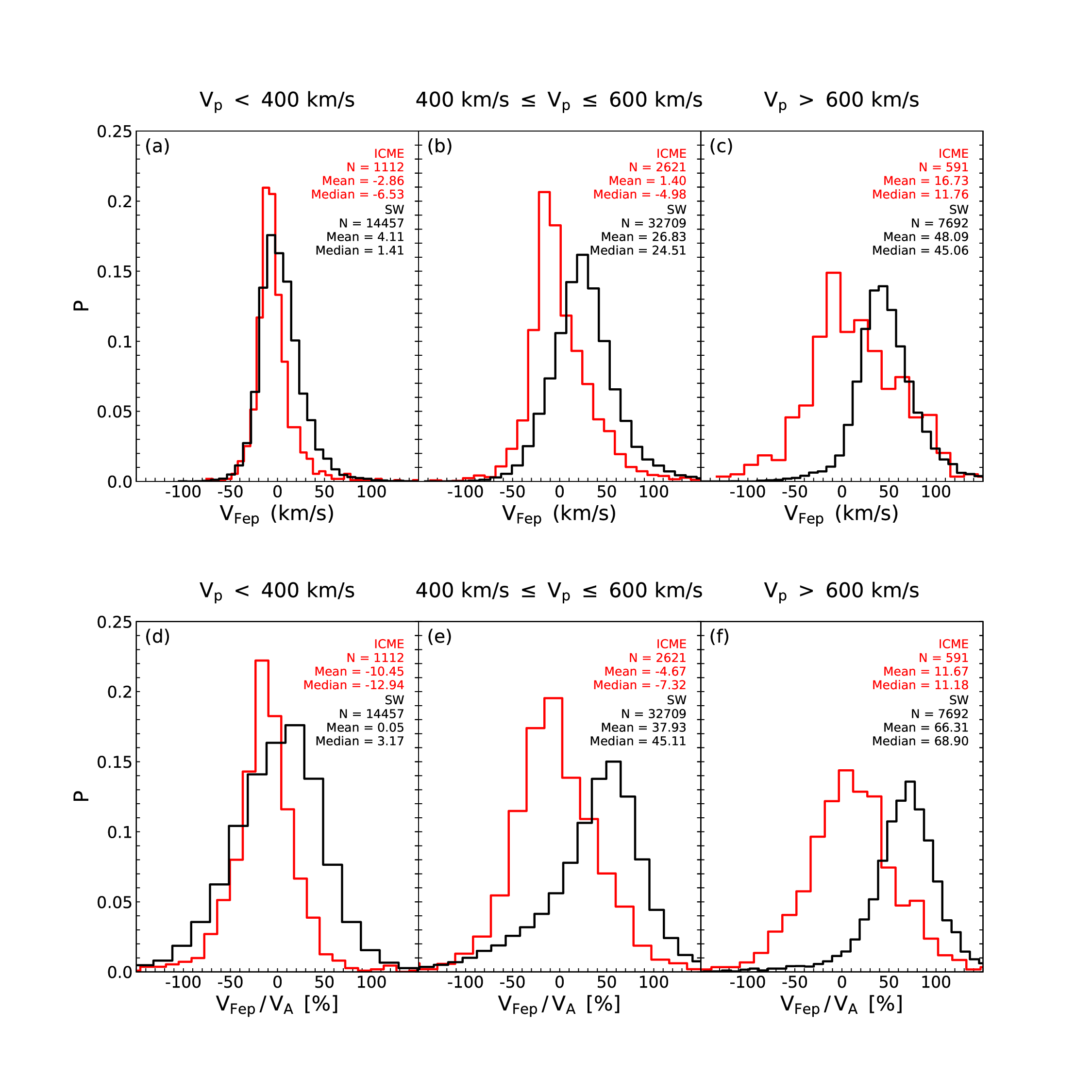} \caption{Same as Figure 1 but for the Fe$^{10+}$-proton differential speed ($V_{Fep}$). \label{Figure 4}}
\end{figure*}

\begin{figure*}[htb!]
\epsscale{1.0} \plotone{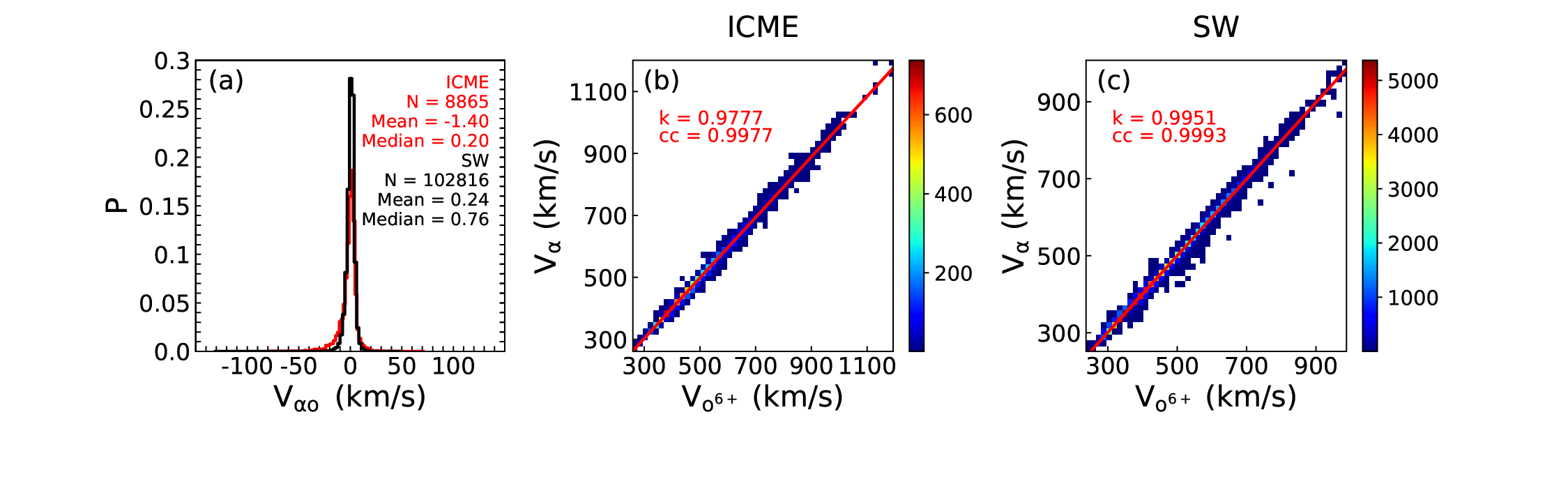} \caption{(a) Probability distributions of the alpha-O$^{6+}$ differential speed ($V_{\alpha o}$) within ICMEs (red) and solar wind (black). (b) and (c) Correlograms of the bulk speeds of alpha particle and O$^{6+}$ within ICMEs and solar wind, respectively. The slope (k) of linear fitting and Pearson correlation coefficient (cc) are denoted in the Panels. \label{Figure 5}}
\end{figure*}

\begin{figure*}[htb!]
\epsscale{1.0} \plotone{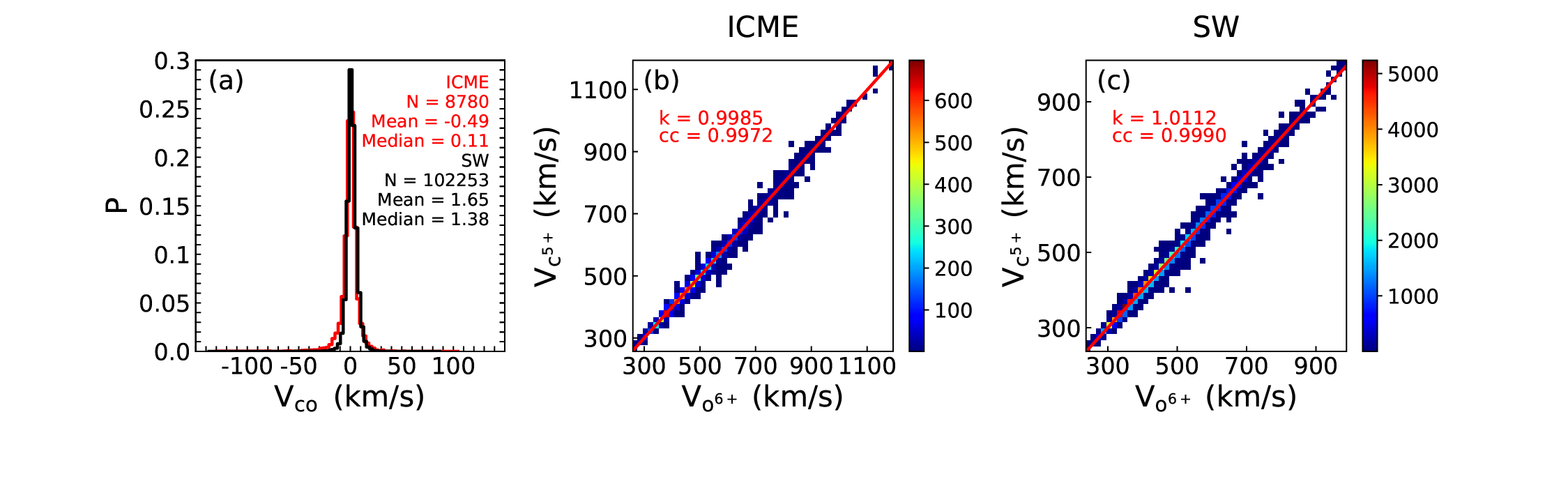} \caption{Same as Figure 5 but for the analysis of C$^{5+}$-O$^{6+}$ differential speed ($V_{co}$). \label{Figure 6}}
\end{figure*}

\begin{figure*}[htb!]
\epsscale{1.0} \plotone{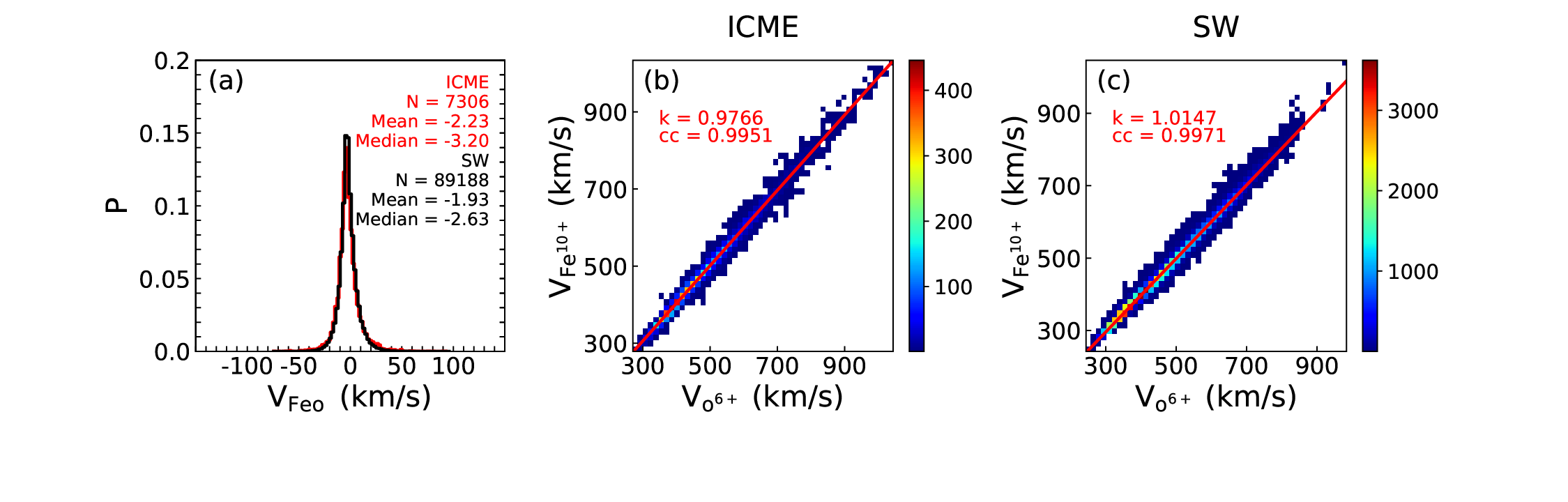} \caption{Same as Figure 5 but for the analysis of Fe$^{10+}$-O$^{6+}$ differential speed ($V_{Feo}$). \label{Figure 7}}
\end{figure*}

\begin{figure*}[htb!]
\epsscale{1.0} \plotone{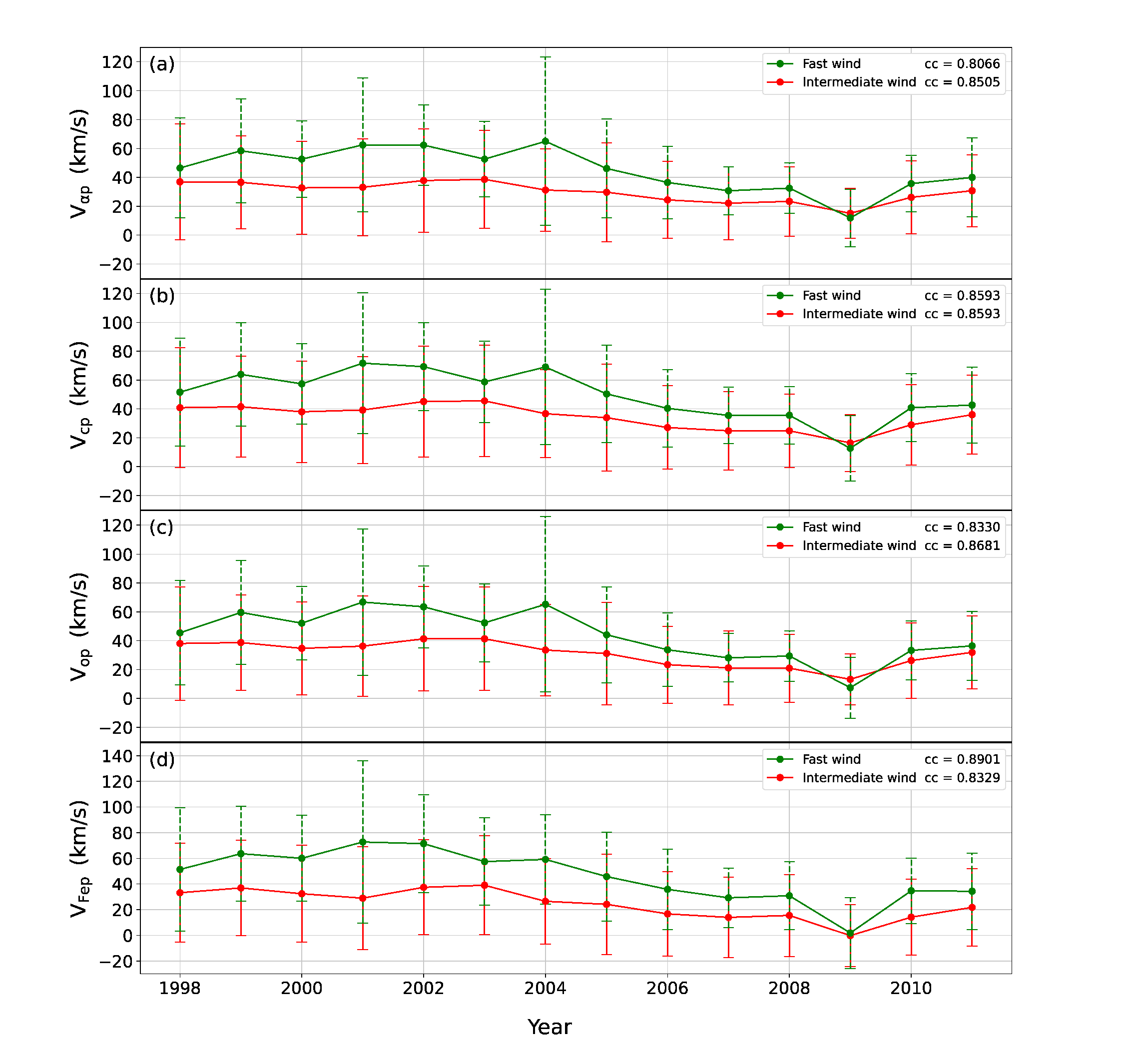} \caption{Solar cycle dependence of ion-proton differential speed within fast (green) and intermediate (red) solar winds from 1998 to 2011. The panels show the yearly variations of $V_{\alpha p}$ (a), $V_{c p}$ (b), $V_{o p}$ (c), and $V_{Fe p}$ (d). The bars indicate the standard deviations in each year. The Spearman correlation coefficients (cc) between the yearly $V_{ip}$ and sunspot numbers are denoted in each panel. \label{Figure 8}}
\end{figure*}

\end{document}